# Excitonic Zeeman Splittings in Colloidal CdSe Quantum Dots Doped with Single Magnetic Impurities

Charles J. Barrows, Rachel Fainblat, and Daniel R. Gamelin[*]

*Department of Chemistry, University of Washington, Seattle, WA 98195-1700*

Email: gamelin@chem.washington.edu

*Unique magneto-optical properties are observed in colloidal quantum dots containing single $Mn^{2+}$ impurities.*

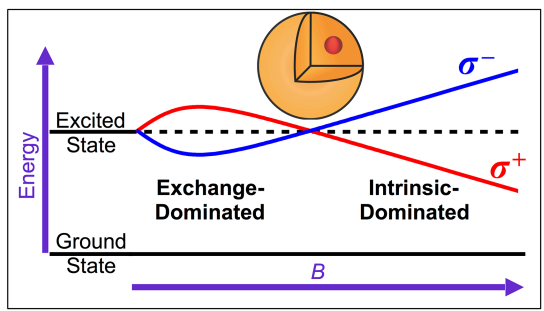

**Abstract.** Doping a semiconductor quantum dot with just a single impurity atom can completely transform its physical properties. Here, we report and analyze the magnetic circular dichroism (MCD) spectra of colloidal CdSe quantum dot samples containing on average fewer than one $Mn^{2+}$ per quantum dot. Even at this sub-single-dopant level, the low-temperature low-field data are dominated by impurity-induced Zeeman splittings caused by dopant–carrier *sp–d* exchange. Unlike in more heavily doped quantum dots, however, the MCD intensity at the first CdSe exciton shows a field-induced sign flip as the field strength is increased, reflecting competition between *sp–d* exchange and the intrinsic Zeeman splittings of comparable magnitude. Most unusually, the competition between these two effects leads to a large apparent shift in the first MCD peak maximum, which we show is attributable to a difference in sign of the intrinsic excitonic *g* factor between the first and second excitons. Finally, the *sp–d* and intrinsic contributions to the excitonic Zeeman splittings each exhibit unique magnetic-field and temperature dependencies, allowing the MCD spectra of undoped, singly doped, and bi-doped quantum dot sub-ensembles to be analyzed.

Incorporation of impurities allows the electrical properties of semiconductors to be tuned, enabling the development of transistors, diodes, and other workhorses of solid-state technology. In addition to electrical effects, impurities can also be used to modify the optical and magnetic properties of semiconductors. Addition of paramagnetic transition metal ions to bulk semiconductors yields so-called "diluted magnetic semiconductors" (DMSs), which combine magnetic and semiconducting properties in a single material.[1,2] In DMSs, dopant-carrier *sp–d* magnetic exchange coupling gives rise to such characteristic magneto-electric and magneto-optical effects as giant excitonic Zeeman splittings,[3] giant Faraday rotation,[1] electrical spin polarization,[4,5] and photo-induced magnetization,[6,7] making these materials promising candidates for applications in spin-electronics and spin-photonics.[8-10]

$Mn^{2+}$-doped CdSe has long served as a model system for investigation of DMSs at the nanoscale.[11,12] Recent advances in nanocrystal diffusion doping have enabled the preparation of high-quality colloidal $Cd_{1-x}Mn_xSe$ QDs with dopant concentrations tunable from 0 to ~30%.[13-15] Due to their smaller volumes, such colloidal DMS QDs can show even stronger *sp–d* exchange coupling than the corresponding bulk materials or epitaxial quantum dots.[16,17] Moreover, their magneto-optical properties can be engineered synthetically by tuning nanocrystal sizes, shapes, compositions, and heterointerfaces to control the spatial overlap between the dopants and the confined semiconductor charge carriers. Such diffusion-doped $Cd_{1-x}Mn_xSe$ QDs have already been used to demonstrate exceedingly large excitonic Zeeman splittings (with $g_{Exc}$ values up to −1180),[18,19] valence-band mixing effects,[20] picosecond spin dynamics in robust excitonic magnetic polarons,[19] and the impacts of $Mn^{2+}$ spin fluctuations.[19,21] Recently, we reported the observation of giant zero-field excitonic exchange splittings in colloidal $Cd_{1-x}Mn_xSe$ QDs containing single $Mn^{2+}$ dopants.[22] Due to the small volumes of these QDs, the largest *sp–d* exchange splittings were almost two orders of magnitude greater than those reported for epitaxial DMS QDs.[23-26] This observation motivates additional spectroscopic characterization of colloidal $Cd_{1-x}Mn_xSe$ QDs in this low-doping regime.

Here, we report MCD studies of ensembles of colloidal $Cd_{1-x}Mn_xSe$ QDs possessing on average fewer than one $Mn^{2+}$ dopant per QD. Low-temperature MCD spectra of these ensembles show a very unusual magnetic-field dependence, including a sign inversion of the first exciton at moderate magnetic fields but no sign inversion of the second exciton. The MCD intensities at the first excitonic transition are shown to reflect the competition between comparable but opposing intrinsic and *sp–d* exchange contributions to the excitonic Zeeman splitting, whereas the intrinsic and *sp–d* exchange contributions add constructively at the second exciton. Analysis of the experimental data yields the MCD spectra of QD sub-ensembles containing exactly one and two $Mn^{2+}$ dopants per QD. Even at just one $Mn^{2+}$ per QD, *sp–d* exchange dominates over the intrinsic Zeeman splittings, but their comparable magnitudes yield anomalous spectral field dependencies, highlighting the unique magneto-optical properties of QDs in this low-doping limit.

$Mn^{2+}$-doped CdSe QDs were prepared by the diffusion-doping method detailed previously[13-15] (see Methods). Figure 1A shows a representative transmission electron microscopy (TEM) image of pseudo-spherical $Cd_{1-x}Mn_xSe$ QDs with diameter $d = 5.1 \pm 0.3$ nm. From inductively coupled plasma atomic emission spectroscopy (ICP-AES), $x = 0.00053 \pm 0.00005$, corresponding to an average of $0.70 \pm 0.07$ $Mn^{2+}$ dopants per nanocrystal. Using these values and a Poisson distribution function (modified to take into account the QD Gaussian size distribution[27]) yields the expected dopant distribution shown in red circles in Figure 1B, in which ~51% of the QDs are undoped, ~34% have exactly 1 $Mn^{2+}$/QD, and only ~12% have 2 $Mn^{2+}$/QD. Independently, the *effective* doping level was also determined by MCD spectroscopy (*vide infra*) and the corresponding distribution (which also takes into account the QD Gaussian size distribution) is shown in blue bars in Figure 1B. Here, the average $Mn^{2+}$ concentration is $0.40 \pm 0.02$/QD ($x = 0.00030 \pm 0.00002$), which corresponds to ~68% of the QDs being undoped, ~26% having exactly 1 $Mn^{2+}$/QD, and ~5% having exactly 2 $Mn^{2+}$/QD. The average doping concentrations determined by the two different methods (ICP-AES and MCD) agree reasonably well and their difference may reflect a small deviation from strictly statistical $Mn^{2+}$ doping in this sample. For the remainder of our analyses we will use the *effective* doping level determined by MCD spectroscopy, because this corresponds to the magneto-optically active $Mn^{2+}$ concentration, but the same conclusions are reached when using the analytical value.

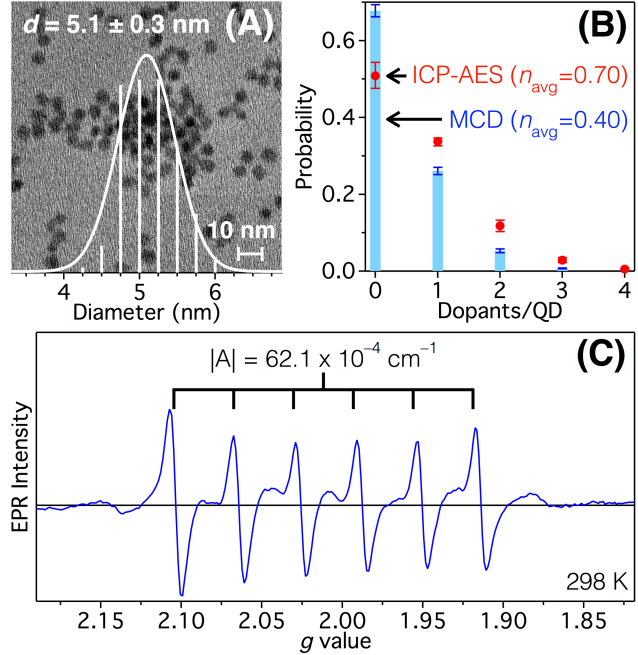

**Figure 1. (A)** TEM image and corresponding size histogram of $d = 5.1 \pm 0.3$ nm ($\sigma = 5.9\%$) $Cd_{1-x}Mn_xSe$ QDs. **(B)** Poisson distribution of dopants taking into account the QD Gaussian size distribution, determined from analysis of the size histogram in panel A and calculated from the average $Mn^{2+}$ concentration determined by ICP-AES (red circles; $n_{avg} = 0.70$ $Mn^{2+}$/QD) or MCD (blue bars; $n_{avg} = 0.40$ $Mn^{2+}$/QD). **(C)** Room-temperature EPR spectrum of these QDs, showing well-resolved $Mn^{2+}$ hyperfine structure ($|A| = 62.1 \times 10^{-4}$ cm$^{-1}$) consistent with the very low $Mn^{2+}$ concentration.

Figure 1C shows a room-temperature EPR spectrum of these QDs. Six hyperfine lines of approximately equal intensity are observed, with a splitting of $|A| = 62.1 \times 10^{-4}$ cm$^{-1}$, consistent with substitutional $Mn^{2+}$ in CdSe.[28] The narrow feature width and absence of broad underlying intensity are both consistent with negligible $Mn^{2+}$–$Mn^{2+}$ dipolar coupling, as anticipated at this low doping concentration.

Figure 2A shows the 1.7 K zero-field electronic absorption spectrum of the QDs from Figure 1, and Figure 2B plots 1.7 K variable-field MCD spectra collected from 0 to 6 T for the same sample. At low magnetic fields, the low-energy leading-edge intensity of the MCD spectrum grows increasingly positive with increasing magnetic field, indicating a negative Zeeman splitting consistent with $Mn^{2+}$-doped CdSe.[12,18,29] This intensity reaches a maximum at ~1.5 T and then decreases and changes sign at higher fields, indicating a positive Zeeman splitting above ~4 T. At the maximum magnetic field (6 T), the MCD spectrum of this sample more closely resembles that of undoped CdSe QDs[30] than of any $Cd_{1-x}Mn_xSe$ QDs reported previously (see ESI).

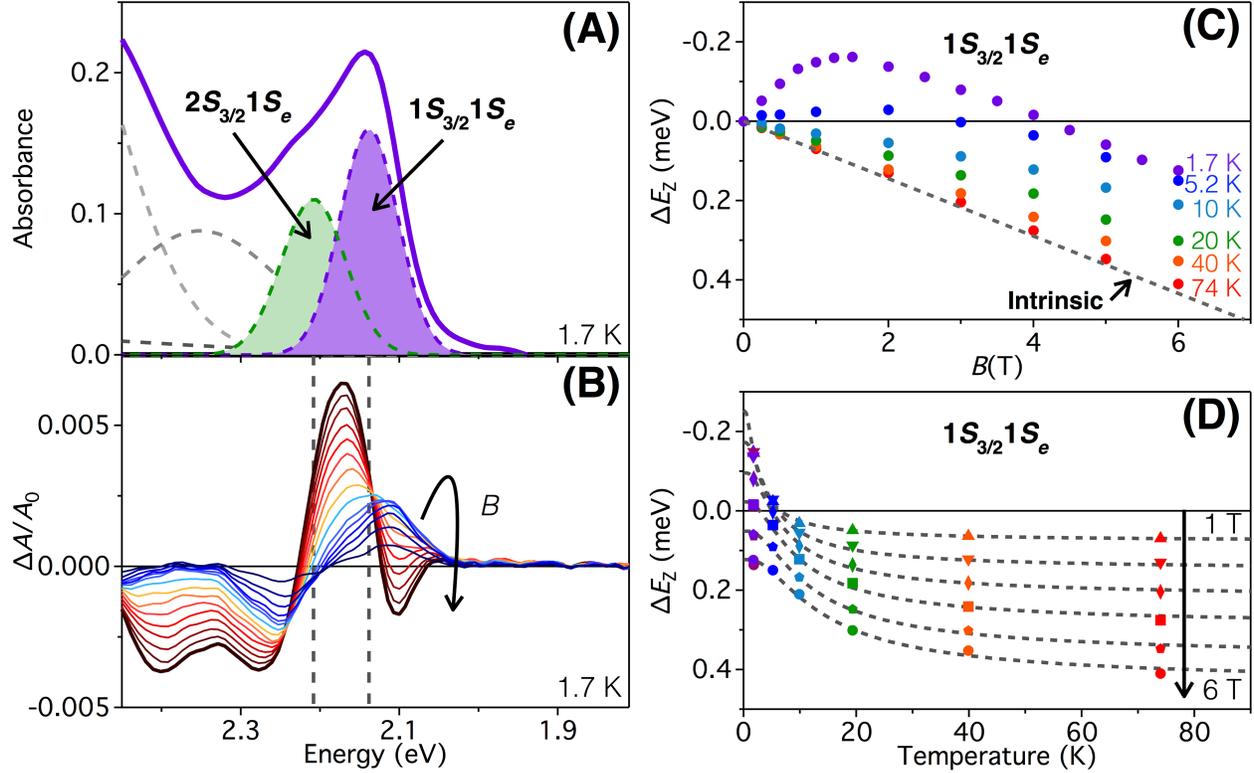

**Figure 2.** **(A)** 1.7 K electronic absorption spectrum (purple line) and multi-peak Gaussian fit (dashed lines) of $d$ = 5.1 nm $Cd_{0.9997}Mn_{0.0003}Se$ QDs, showing the first two allowed excitonic transitions (filled curves). **(B)** Corresponding variable-field (0–6 T; blue to red lines) MCD data for the same QDs at 1.7 K. The arrow indicates the direction of increasing magnetic field. The vertical dashed lines indicate the average energies of the first two excitonic transitions. **(C)** Field-dependent Zeeman splitting energies of the first exciton, determined from analysis of these variable-temperature, variable-field MCD spectra. (Spectra collected at temperatures above 1.7 K are included in the ESI.) The intrinsic component (which is temperature-independent) is plotted as a grey dashed line. **(D)** Temperature dependence of the first excitonic Zeeman splitting from 1 T (upward triangles) to 6 T (circles). The grey dashed lines represent a global fit to the data as described in the text, and correspond to $x_{eff}$ = 0.00030 (0.40 $Mn^{2+}$/QD).

Excitonic Zeeman splittings ($\Delta E_Z$) of the $1S_{3/2}1S_e$ exciton were calculated from the absorption and MCD data of Figure 2A,B using Equation 1,[12,18] where $\Delta A'$ corresponds to the maximum amplitude of the lowest-energy leading-edge MCD feature and $A_0$ and $\sigma$ are the height and Gaussian width of the absorption peak, respectively, determined by multi-peak Gaussian fitting using fixed heights and widths across the magnetic field range, according to the rigid-shift approximation.[12,31] Figure 2C plots $\Delta E_Z$ as a function of magnetic field and temperature, showing the inversion of $\Delta A'/A_0$ with magnetic field seen in the raw 1.7 K spectra of Figure 2B. Similar results are obtained from the 5.2 K data, but the sign inversion occurs at a smaller field. At 10 K and above, $\Delta E_Z$ is positive at every magnetic field. Absorption and MCD spectra collected at temperatures above 1.7 K are included in the ESI.

$$\Delta E_Z = \frac{\sqrt{2e}}{2}\sigma \frac{\Delta A'}{A_0} \qquad (1)$$

To extract an effective $Mn^{2+}$ concentration, these Zeeman splitting data were replotted *vs* temperature (Figure 2D) and analyzed. Following the approach of ref. 32, the data in Figure 2D were fit to Equation 2, where $x_{eff}$ is the effective $Mn^{2+}$ concentration, $N_0(\alpha-\beta)$ describes the bulk *sp–d* exchange (1.5 eV),[33] and $\langle S_Z \rangle$ is the spin expectation value of $Mn^{2+}$, which follows Brillouin behavior for a spin-only $S = -5/2$ ground state (Equation 3; $g_{Mn} = 2.0042$ and $T$ = experimental temperature). Equation 2 describes $\Delta E_Z$ as the sum of intrinsic ($\Delta E_{Int}$) and *sp–d* exchange ($\Delta E_{sp-d}$) terms, where $\Delta E_{Int}$ is temperature-independent and scales linearly with field, $g_{Int}$ is the intrinsic excitonic $g$ value, $\mu_B$ is the Bohr magneton, and $B$ is the magnetic field. The dashed lines in Figure 2D show a global best fit to these data, yielding $x_{eff} = 0.00030 \pm 0.00002$, which corresponds to 0.40 $Mn^{2+}$/QD and the doping statistics plotted in Figure 1B.

$$\Delta E_Z = g_{Int}\mu_B B + x_{eff} N_0(\alpha-\beta)\langle S_Z \rangle \qquad (2a)$$

$$= \Delta E_{Int} + \Delta E_{sp-d} \qquad (2b)$$

$$S_Z = \frac{2S+1}{2}\coth\left(\frac{2S+1}{2}\cdot\frac{g_{Mn}\mu_B B}{k_B T}\right) - \frac{1}{2}\coth\left(\frac{g_{Mn}\mu_B B}{2k_B T}\right) \qquad (3)$$

All MCD spectra of other $Cd_{1-x}Mn_xSe$ QDs reported previously have been overwhelmingly dominated by the *sp–d* exchange terms. In contrast, the MCD spectra in Figure 2B show clear evidence of both intrinsic and *sp–d* exchange contributions simultaneously. Whereas $\Delta E_{sp-d}$ becomes field-independent when all of the $Mn^{2+}$ spins have been aligned by the magnetic field ($\langle S_Z \rangle = -5/2$), $\Delta E_{Int}$ continues to increase linearly with $B$ (Equation 2a). In this regime, *e.g.*, above ~2.5 T at 1.7 K in Figure 2C, the intrinsic Zeeman contribution to the MCD spectra of Figure 2B can be isolated according to Equation 4, where $(\Delta A/A_0)_B$ is the MCD spectrum at magnetic field $B$.

$$(\Delta A/A_0)_{Int,1} = (\Delta A/A_0)_6 - (\Delta A/A_0)_5 \qquad (4)$$

Scaling $(\Delta A/A_0)_{Int,1}$ to experimental magnetic fields from 0 to 6 T yields the spectra shown in Figure 3A. Analysis of these spectra gives $g_{Int} = +1.3$, which agrees well with literature values for CdSe QDs of this size.[30] Subtraction of these spectra from the data in Figure 2B yields intensities attributable solely to $sp$–$d$ exchange (Figure 3B). As expected, the low-energy leading edge of these spectra has the opposite sign compared to the intrinsic spectra, and its amplitude follows the $S = -5/2$ Brillouin curve.

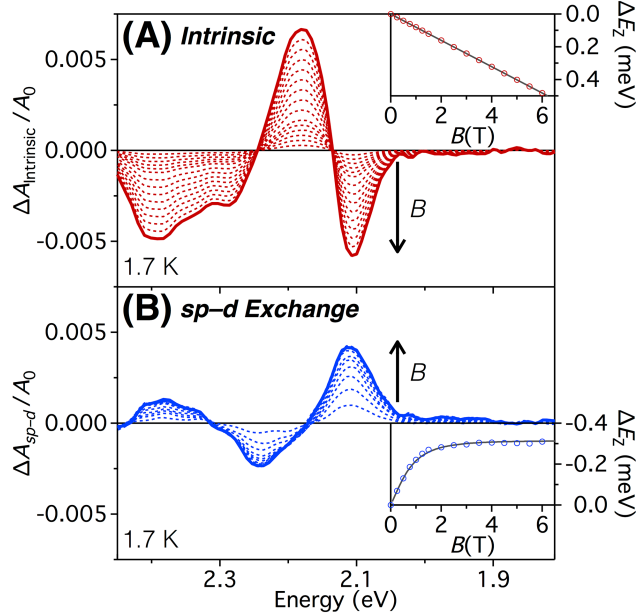

**Figure 3. (A)** Intrinsic component of the variable-field MCD spectra from Figure 2B, deconvolved according to Equation 4. **(B)** $sp$–$d$ exchange component of the MCD spectra from Figure 2B, deconvolved by subtracting the spectra of Figure 3A from the spectra of Figure 2B. Arrows indicate the direction of increasing magnetic field (0–6 T). Insets: Field dependence of the intrinsic (top) and $sp$–$d$ exchange (bottom) contributions to the first exciton's Zeeman splitting. The $sp$–$d$ exchange term was fit to an $S = -5/2$ Brillouin curve (grey line).

The MCD data in Figure 3 allow reconstruction of MCD spectra of QD *ensembles* doped with specific integer numbers of $Mn^{2+}$ ions (rather than with a Poisson distribution of dopants) using Equation 5, where $n$ indicates the integer number of dopants per QD. Equation 5 assumes that the $sp$–$d$ exchange term $(\Delta A(B)/A_0)_{sp-d}$ scales linearly with $n$, which has previously been demonstrated for such QDs[14] in the limit of low *x*, where $Mn^{2+}$–$Mn^{2+}$ interactions are negligible. The resulting reconstructed spectra are shown in Figure 4A,B. At 2 $Mn^{2+}$/QD (Figure 4A), the MCD spectra resemble the "typical" data reported for other $Cd_{1-x}Mn_xSe$ QDs, where *sp–d* exchange dominates and $\Delta A/A_0$ is positive at the leading edge of the first exciton.[13,29] At 1 $Mn^{2+}$/QD (Figure 4B), the MCD spectra also display a positive leading edge at all experimental fields, but $\Delta A'/A_0$ clearly turns over at ~2 T, and the peak maximum appears to blue-shift by ~40 meV over the full field range. Figure 4C plots $\Delta E_Z$ of the $1S_{3/2}1S_e$ exciton, extracted from each of the reconstructed spectra in Figure 4A–B, as well as $\Delta E_{Int}$ from Figure 3A. At all experimental fields, $\Delta E_Z < 0$ for the QDs with quantized dopants and $\Delta E_Z > 0$ for undoped QDs.

$$\left(\frac{\Delta A(B)}{A_0}\right)_n = \left(\frac{\Delta A(B)}{A_0}\right)_{Int} + \frac{n}{n_{avg}}\left(\frac{\Delta A(B)}{A_0}\right)_{sp-d} \tag{5}$$

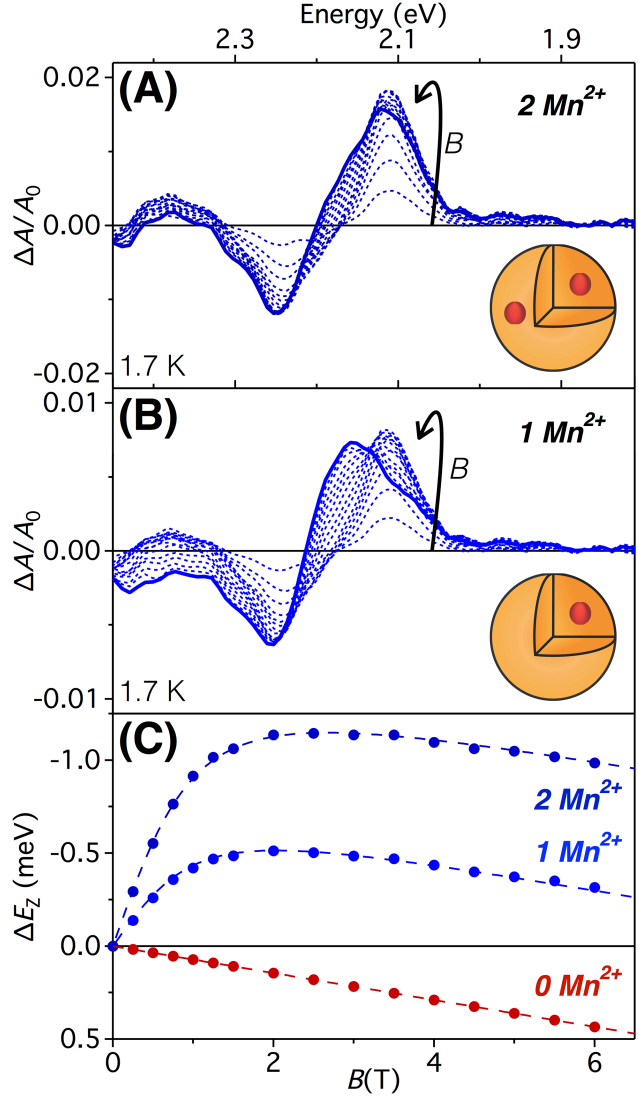

**Figure 4.** Variable-field MCD spectra measured at 1.7 K, deconvolved from the data in Figure 2B according to Equation 5 for $Cd_{1-x}Mn_xSe$ QDs containing exactly 2 **(A)** or exactly 1 **(B)** $Mn^{2+}$ per QD. The arrows indicate the direction of increasing magnetic field from 0–6 T. **(C)** Corresponding Zeeman splittings of the $1S_{3/2}1S_e$ transition in each of these data sets, and also for undoped CdSe QDs from Figure 3A.

It is remarkable that the *sp–d* exchange dominates over the intrinsic excitonic Zeeman splitting after adding only 1 $Mn^{2+}$ in a $d = 5.1$ nm CdSe QD. This result is particularly significant because the data here reflect the full distribution of radial $Mn^{2+}$ positions within the ensemble of QDs, meaning the *sp–d* exchange strength is sufficiently large for just 1 $Mn^{2+}$ to cause a giant magneto-optical response even when that $Mn^{2+}$ is located at the average cation radius instead of near the QD center. Although an MCD sign inversion was previously observed for CdSe clusters doped with 1 $Mn^{2+}$ per cluster,[34,35] the effective dopant concentration in these QDs is approximately two orders of magnitude smaller than in those clusters. In self-assembled $Cd_{1-x}Mn_xSe$ QDs, inversion of the excitonic Zeeman splitting has been predicted for effective dopant concentrations above $x \approx 0.001$, but only tested for samples that contained nominal dopant concentrations above 1%.[36] These results are most reminiscent of the non-monotonic field dependence observed in magneto-reflectivity measurements of $Cd_{1-x}Mn_xTe$ quantum wells with extremely small values of $x$.[37] We note that the data for 1 $Mn^{2+}$/QD suggest that the average QD's first-exciton exchange splitting is 0.64 meV even at zero applied magnetic field. Interestingly, this value is smaller than the exciton linewidths observed in the single-particle photoluminescence spectra attributed to undoped NCs within similar $Cd_{1-x}Mn_xSe$ QD ensembles (~6 meV at 5 K).[22] These results thus confirm the hypothesis presented in ref. 22 that single-QDs showing much larger zero-field excitonic *sp–d* exchange splittings (approaching ~80 meV) must have their $Mn^{2+}$ ions significantly closer to their centers than the average cation position, where dopant-exciton spatial overlap (and hence also the *sp–d* exchange coupling strength) is maximized. We estimate that the *sp–d* exchange energy is ~50 times greater for a $Mn^{2+}$ at the exact center of a $d = 5.1$ nm CdSe QD than for a $Mn^{2+}$ at the average distance from the QD center,[16,22] consistent with this conclusion.

Finally, we analyze the anomalous spectral evolution with magnetic field shown in Figures 2B and 4B. This behavior appears to be unique to the case of ≤1 $Mn^{2+}$/QD (avg.), where $\Delta E_{Int}$ and $\Delta E_{sp-d}$ are most comparable in magnitude. Analysis shows that this unusual behavior derives from a change in the relative *signs* of $\Delta E_{Int}$ and $\Delta E_{sp-d}$ between the first ($1S_{3/2}1S_e$) and second ($2S_{3/2}1S_e$) excitonic transitions. Figure 5A,B summarizes the contrast between the first and second excitons of the full ensemble (0.4 $Mn^{2+}$/QD) by plotting the field-dependence of $\Delta E_{Int}$ and $\Delta E_{sp-d}$ for each transition. The sums of these individual contributions yield the black dashed lines, which overlay with the total Zeeman splittings calculated from the raw spectra of Figure 2B (black circles and triangles). Figure 5C–E re-plots the 2, 4, and 6 T MCD spectra of Figure 2B, respectively, along with the Gaussian fit results for the first two excitonic transitions. At low fields, $\Delta E_Z$ is dominated by $\Delta E_{sp-d}$ which is negative for both the $1S_{3/2}1S_e$ and $2S_{3/2}1S_e$ transitions, resulting in MCD spectra like those of Figure 5C. As the field increases, $\Delta E_{sp-d}$ saturates and $\Delta E_{Int}$ begins to rival it in magnitude. $\Delta E_{Int}$ is positive for the $1S_{3/2}1S_e$ exciton, causing it to oppose and diminish this exciton's MCD intensity, but $\Delta E_{Int}$ is negative for the $2S_{3/2}1S_e$ exciton,[20] causing the MCD intensity of this exciton to continue growing with increasing field. Near 4 T, the magnitudes of $\Delta E_{Int}$ and $\Delta E_{sp-d}$ for the $1S_{3/2}1S_e$ exciton are nearly identical, leading to complete cancellation of this exciton's MCD intensity. The experimental MCD spectrum at 4 T (Figure 5D) therefore shows only higher-energy transitions, beginning with the $2S_{3/2}1S_e$ exciton. At higher fields (Figure 5E), $\Delta E_Z$ is dominated by $\Delta E_{Int}$. This contrast between opposing *vs* additive combinations of intrinsic and *sp–d* contributions is only observed in this extremely low doping limit, and explains the anomalous MCD spectral characteristics and intensity inversion in the MCD spectra of Figure 2B.

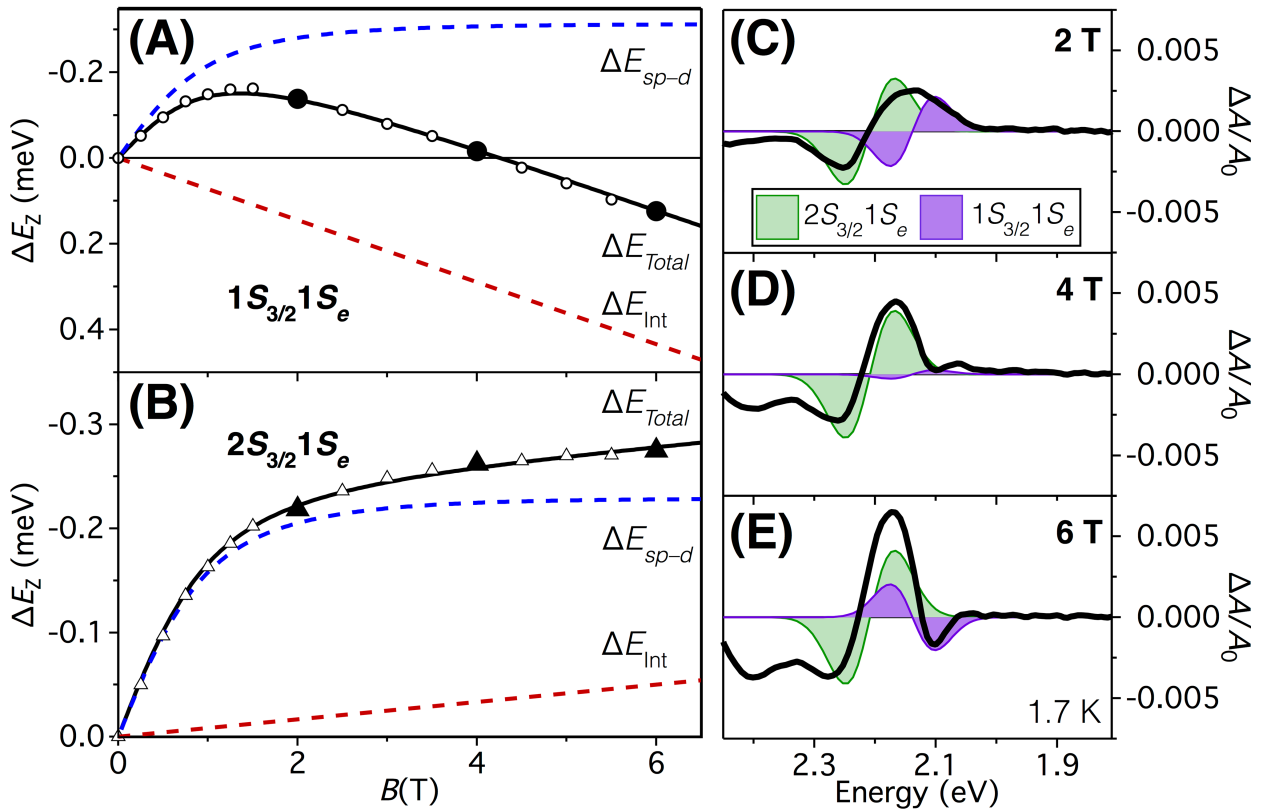

**Figure 5.** **(A–B)** Total Zeeman splittings (black) of the $1S_{3/2}1S_e$ (**A**; circles) and $2S_{3/2}1S_e$ (**B**; triangles) excitonic transitions as determined from analysis of the magnetic field dependent MCD spectra of $Cd_{0.9997}Mn_{0.0003}Se$ QDs at 1.7 K, divided into their intrinsic (red) and *sp–d* exchange (blue) components. Closed symbols correspond to the three sets of MCD spectra plotted in panels C–E. **(C–E)** MCD spectra (black) at 1.7 K and 2 T **(C)**, 4 T **(D)**, and 6 T **(E)**, along with their respective contributions from the $1S_{3/2}1S_e$ (purple) and $2S_{3/2}1S_e$ (green) transitions.

In summary, MCD spectra of colloidal CdSe QDs doped with trace $Mn^{2+}$ reflect the rich relationship between intrinsic and *sp–d* exchange contributions to the excitonic Zeeman splittings of these materials. Opposing and additive contributions of these terms to $\Delta E_Z$ in the $1S_{3/2}1S_e$ and $2S_{3/2}1S_e$ transitions, respectively, generate abnormal bandshapes and an intensity turnover in the variable-field MCD spectra of these QDs. Spectral deconvolution allows reconstruction of the MCD spectra for QD ensembles possessing exactly 1 and 2 $Mn^{2+}$/QD, which are both still dominated by *sp–d* exchange even when the average dopant is not located in the QD center. These data reveal unique spectroscopic properties in ensembles of colloidal QDs containing exactly one $Mn^{2+}$ per QD, despite radial doping distributions, and they highlight the ability of even single dopants to transform the physical properties of quantum-confined semiconductor nanostructures.

**Experimental Methods**

**Synthesis.** $Cd_{1-x}Mn_xSe$ nanocrystals were prepared by diffusion doping in the $Se^{2-}$-limited regime according to the methods described in refs 13,14. After doping, the QDs were cooled to room temperature and washed by repeated suspensions in toluene and flocculation with ethanol.

**Physical characterization.** Atomic concentrations were determined by analysis of dried nanocrystals digested in ultrapure nitric acid (EMD Chemicals) using inductively coupled plasma atomic emission spectrometry (ICP-AES; Perkin-Elmer). Room-temperature electron paramagnetic resonance (EPR) experiments were performed on colloidal toluene suspensions of nanocrystals using an X-band Bruker EMX spectrometer. Low-temperature absorption and magnetic circular dichroism (MCD) spectra were collected on nanocrystal films prepared by depositing a toluene suspension of nanocrystals between quartz disks. The films were placed in a superconducting magneto-optical cryostat (Cryo-Industries SMC-1659 OVT) oriented in the Faraday configuration. At helium temperature, the sample was screened for depolarization by matching the CD spectra of a chiral molecule placed before and after the sample. Depolarization of the thin film was <5 %. Electronic absorption and MCD spectra were collected simultaneously using an Aviv 40DS spectropolarimeter. The differential absorption collected in the MCD experiment is reported as $\Delta A = A_L - A_R$, where $A_L$ and $A_R$ refer to the absorption of left and right circularly polarized photons in the sign convention of Piepho and Schatz.[12,31] From these data, values of $\Delta E_{Zeeman}$ and $g_{Exc}$ can be obtained.[12,13,18]

**Acknowledgments.** This research was funded by the US National Science Foundation (DMR-1505901 to D.R.G.). R.F. was supported by the German Academic Exchange Service (DAAD) with funds from the German Federal Ministry of Education and Research (BMBF) and the People Programme (Marie Curie Actions) of the European Union's Seventh Framework Programme (FP7/2007-2013) under REA grant agreement n° 605728 (P.R.I.M.E. – Postdoctoral Researchers International Mobility Experience). The authors thank Mr. Michael De Siena for performing TEM measurements.

**Electronic Supplementary Information (ESI) available:** Additional absorption and MCD spectra, Zeeman-splitting analysis, and a table of *g* values. See DOI: 10.1039/x0xx00000x.